\documentclass{PoS}
\usepackage{graphicx}
\usepackage{cases}

\PoS{PoS(LAT2005)047}

\title{Staggered Lattice Artifacts in 3-Flavor Heavy Baryon Chiral Perturbation Theory }

\ShortTitle{Staggered Heavy Baryon Chiral Perturbation Theory }

\author{\speaker{Jon A. Bailey} and C. Bernard\\
        Washington University in St. Louis\\
        E-mail: \email{jabailey@wustl.edu}, \email{cb@lump.wustl.edu}}

\abstract{Motivated by simulation results for octet and decuplet baryon masses using 2+1 flavors of light staggered quarks, we are incorporating staggered lattice artifacts into heavy baryon chiral perturbation theory, calculating the masses of various staggered baryons, and studying the connection between the staggered baryons of the chiral theory and the staggered baryons of simulations.
We present ${\cal O}(m_q^{3/2})$ loop contributions to the masses of several staggered nucleons and discuss interpolating fields that create these states.}

\FullConference{XXIIIrd International Symposium on Lattice Field Theory\\
		 25-30 July 2005\\
		 Trinity College, Dublin, Ireland}

\begin{document}

\section{\label{intro}Introduction}

The development of staggered chiral perturbation theory (S$\chi$PT) for light and heavy-light mesons \cite{LEE-SHARPE,1,3} has helped improve the precision of calculations of light quark masses, meson masses, and meson decay constants \cite{4,CHARMED}.  In contrast, staggered calculations of baryon masses have used continuum chiral forms fit to continuum-extrapolated lattice results \cite{11}.  While rigorous in principle, it is difficult to control the continuum extrapolation without a detailed understanding of how taste violating effects enter at fixed lattice spacing.  In addition, performing a continuum extrapolation before the chiral extrapolation reduces the amount of data available and obscures the correlations in partially quenched data.  Incorporating staggered artifacts into heavy baryon chiral perturbation theory (HB$\chi$PT) allows us to quantify the taste violations and investigate the possibility of precision calculations of baryonic quantities.  In addition, this approach makes evident the taste splitting in the spectra of staggered baryon interpolating fields, which in turn has implications for fits to staggered baryon correlators.  In the discussion that follows, we focus on staggered baryons that are degenerate with the nucleon in the continuum limit.




\section{\label{id}Identifying nucleons in the staggered baryon spectrum}


Because of taste symmetry, the staggered baryon spectrum contains more states than the physical spectrum.  Incorporating taste, the flavor $SU(3)$ becomes an $SU(12)_{\rm val}$ for the fermionic valence quarks even if the $4^{\rm th}$ root is used to eliminate the taste degree of freedom of the sea quarks.
We find that staggered HB$\chi$PT contains 572 spin-$1\over 2$ baryons corresponding to the ground-state octet and 364 spin-$3\over 2$ baryons corresponding to the decuplet.
Decomposing the {\bf 572} under the flavor-taste $SU(3)_F\times SU(4)_T\subset SU(12)_{\rm val}$,
\begin{equation}
{\bf 572_M}\rightarrow{\bf (10_S,\ 20_M)\oplus(8_M,\ 20_S)\oplus(8_M,\ 20_M)\oplus(8_M,\ \bar 4_A)\oplus(1_A,\ 20_M)},\label{mix}
\end{equation}
where the subscripts refer to the symmetries of the multiplets (symmetric, mixed, or antisymmetric).
Taking the isospin and continuum limits, the ${\bf (10_S,\ 20_M)}$ and ${\bf (8_M,\ 20_S)}$ contain a total of 120 baryons that are degenerate with the nucleon \cite{PAPER}.  At nonzero lattice spacing, discretization effects split the staggered nucleons of the ${\bf (10_S,\ 20_M)}$ into four lattice irreducible representations (irreps) and those of the ${\bf (8_M,\ 20_S)}$ into three lattice irreps.  The decomposition of the corresponding spin-taste $SU(2)_S\times SU(4)_T$ irreps under the geometrical time slice group (GTS) of the lattice \cite{17} reflects this splitting:
\begin{eqnarray}
{\bf ({\rm 1\over 2},\ 20_M)}&\rightarrow&{\bf 8}\oplus{\bf 8}\oplus{\bf 8}\oplus{\bf 16}\label{20M to lat}  \label{eq:20M} \\
{\bf ({\rm 1\over 2},\ 20_S)}&\rightarrow&{\bf 8\oplus 16\oplus 16}.\label{eq:20S}
\end{eqnarray}
Therefore, there are seven nondegenerate staggered nucleons.  
Because of flavor isospin symmetry, each of the nucleons in (\ref{eq:20M}) is
replicated four times, and each one in (\ref{eq:20S}) is replicated twice.

\section{\label{loops}Leading loops for masses of ${\bf (10_S,\ 20_M)}$ nucleons}

To ${\cal O}(m_q^{3/2})$ in the staggered chiral expansion, the masses are given by
$m_N = m_0 + {\rm Tree}(m_q) + {\rm Tree}(a^2) + {\rm Loop}{\left(\frac{1}{2}\right)} + {\rm Loop}{\left({3\over 2}\right)}$,
where $m_0$ is the average octet mass in the chiral and continuum limits, ${\rm Tree}(m_q)$ denotes tree-level terms linear in $m_q$, ${\rm Tree}(a^2)$ denotes tree-level terms linear in $a^2$, and ${\rm Loop}(s)$ is the sum of sunset diagrams with virtual spin-$s$ baryons; the loops enter at ${\cal O}(m_q^{3/2})$ and include chiral logarithms proportional to $\Delta\,m_\phi^2\ln\>m_\phi$, where $\Delta$ is the average octet-decuplet mass difference in the chiral and continuum limits, and $m_{\phi}$ is a generic meson mass.  To date we have calculated ${\rm Tree}(m_q)$ and the loops; below we also include estimates for the contributions ${\rm tree}(a^2)\subset {\rm Tree}(a^2)$ that are required for renormalizing connected ${\rm Loop}({3\over 2})$ terms.  Although the terms ${\rm Tree}(a^2)$ generally break the symmetry down to the lattice $\Gamma_4\rtimes SW_{4,\,{\rm diag}}$~\cite{LEE-SHARPE}, all the contributions calculated to date respect the larger remnant taste $\Gamma_4\rtimes SO(4)_T\subset SU(4)_T$~\cite{LEE-SHARPE,1}.  To calculate the remaining ${\rm Tree}(a^2)$ terms, one must map the necessary Symanzik operators into the chiral theory; this task is in progress \cite{PAPER}.  

The leading order heavy baryon Lagrangian is the same as for a continuum $SU(12)_L\times SU(12)_R$ theory; therefore, at one loop taste violations enter baryon masses only through meson propagators.  
The loops at this order are generated by inserting connected and disconnected (hairpin) meson propagators into the vertices of the leading order heavy baryon Lagrangian; respectively denoting these contributions $\sigma\,^{\rm conn}_{s}$ and $\sigma\,^{\rm disc}_{s}$ and adopting the notation of Ref. \cite{CHEN-SAVAGE} for the low-energy parameters,
\begin{eqnarray*}
{\rm Loop}{\left(\frac{1}{2}\right)}=-\frac{1}{48\pi f^2}\>[\,\sigma\,^{\rm conn}_{1/2}+\sigma\,^{\rm disc}_{1/2}\,],&\qquad&{\rm Loop}{\left(\frac{3}{2}\right)}=\left(\frac{C}{8\pi f}\right)^2[\,\sigma\,^{\rm conn}_{3/2}+\sigma\,^{\rm disc}_{3/2}\,].
\end{eqnarray*}

Now we define some notation.
Let $\alpha$ be meson taste; then $\alpha \in \{I,\ \mu,\ \mu\nu\ (\mu < \nu),\ \mu5,\ 5\}$.  Let $t$ be meson taste irrep; then $t \in \{I,\ V,\ T,\ A,\ P\}$.  Let $n_t$ be the number of tastes belonging to the irrep $t$, and let $s_t$ be the set of meson tastes in the irrep $t$, so that\enskip $s_I = \{I\}$,\enskip$s_V = \{\mu\}$,\enskip$s_T = \{\mu\nu\ (\mu < \nu)\}$, and so on. 
For baryons in the ${\bf (10_S,\ 20_M)}$, we define the taste factors
\begin{subnumcases}{\label{R}{\cal R}\,^{\alpha}_{abc} \equiv }
(2/3)[\,(\xi_{aa}^{\alpha})^2+5\,\xi_{ab}^{\alpha}\xi_{ba}^{\alpha}-4\,\xi_{aa}^{\alpha}\xi_{bb}^{\alpha}\,] & for taste N's\nonumber\\
(2/3)(\,\xi_{aa}^{\alpha}\xi_{bb}^{\alpha}+\xi_{ab}^{\alpha}\xi_{ba}^{\alpha}\,)+(1/3)[\,(\,5\,\xi_{ac}^{\alpha}\xi_{ca}^{\alpha}-4\,\xi_{aa}^{\alpha}\xi_{cc}^{\alpha}\,)+(\,a\,\rightarrow\,b\,)\,] & for taste $\Sigma$'s\nonumber\\
-2\,(\,\xi_{aa}^{\alpha}\xi_{bb}^{\alpha}-\xi_{ab}^{\alpha}\xi_{ba}^{\alpha}\,)+[\,\xi_{ac}^{\alpha}\xi_{ca}^{\alpha}+(\,a\,\rightarrow\,b\,)\,] & for taste $\Lambda$'s\nonumber,
\end{subnumcases}
\begin{subnumcases}{\label{S}{\cal S}\,^{\alpha}_{abc} \equiv }
G_+(\xi_{aa}^{\alpha})^2-H_+\xi_{ab}^{\alpha}\xi_{ba}^{\alpha}+J_+\xi_{aa}^{\alpha}\xi_{bb}^{\alpha}\nonumber\\
G_{\pm}\,(\,\xi_{aa}^{\alpha}\xi_{bb}^{\alpha}\pm\xi_{ab}^{\alpha}\xi_{ba}^{\alpha}\,)-(H_{\pm}/2)(\,\xi_{ac}^{\alpha}\xi_{ca}^{\alpha}+\xi_{bc}^{\alpha}\xi_{cb}^{\alpha}\,)+(J_{\pm}/2)(\,\xi_{aa}^{\alpha}\xi_{cc}^{\alpha}+\xi_{bb}^{\alpha}\xi_{cc}^{\alpha}\,),\nonumber
\end{subnumcases}
\begin{center}
\begin{minipage}[t]{0.45\textwidth}
\begin{eqnarray*}
G_+&\equiv&4(3F^2-D^2)\\
H_+&\equiv&4[(3F-D)^2-6F^2]\\
J_+&\equiv&24F(F-D)
\end{eqnarray*}
\end{minipage}%
\hspace{0.3cm}
\begin{minipage}[t]{0.45\textwidth}
\begin{eqnarray*}
G_-&\equiv&(4/3)(9F^2+D^2-12DF)\\
H_-&\equiv&-(4/3)(9F^2-5D^2+6DF)\\
J_-&\equiv&(8/3)(9F^2-2D^2-3DF),
\end{eqnarray*}
\end{minipage}%
\end{center}
where $a,~b,$ and $c$ are quark taste indices, $a,~b,~c \in \{1,\ 2,\ 3,\ 4\}$, and $\xi^{\alpha}$ is a taste matrix.
For 3-taste baryons, the indices $abc$ can equal only the permutations $abc=123$, $124$, $341$, and $342$.  The first row of ${\cal R}\,^{\alpha}_{abc}$ applies for taste nucleons, whose taste structure is isomorphic to the flavor structure of the continuum nucleon.  The second row applies for taste sigmas, and the third, for taste lambdas.  Likewise, the first row of ${\cal S}\,^{\alpha}_{abc}$ applies for taste nucleons; in the second row of ${\cal S}\,^{\alpha}_{abc}$, the upper sign ($+$) applies for taste sigmas and the lower ($-$), for taste lambdas.
Let ${\cal R}\,^t_{abc} \equiv \sum _{\alpha \in s_t}{\cal R}\,^{\alpha}_{abc}$,
and likewise for ${\cal S}$.
We will also need the parameter $K\equiv(3F-D)^2+4D^2$.
Let $r \in \{V,\ A\}$ (cf. the range of $t$ above).  
Defining the sets of indices $L_r\equiv \{\pi_r,\ \eta_r,\ \eta'_r\}$ facilitates summing over the residues $R_l\equiv-R_l^{[3,1]}(\{m_{l'}|l'\in L_r\};\{m_{S_r}\})$~\cite{1}, where $l\in L_r$.  Finally, in ${\rm Loop}({3\over 2})$ terms we use the function ${\cal F}(M)\equiv-{1\over 3}\,F(M,\Delta,\mu)$~\cite{CHEN-SAVAGE}.
Then for the masses of the full ($m_{\rm val}=m_{\rm sea}$) ${\bf (10_S,\ 20_M)}$ nucleons in the isospin limit,\footnote{In the valence isospin limit, the meson taste singlet contributions to $\sigma\,^{\rm disc}_{3/2}$ vanish for both ${\bf (10_S,\ 20_M)}$ and ${\bf (8_M,\ 20_S)}$ nucleons.}
\begin{eqnarray*}
\sigma\,^{\rm conn}_{1/2}&\equiv&\frac{1}{4}\sum _t\left[\,(\,K\,n_t+{\cal S}\,^t_{abc}\,)\,m_{\pi_t}^3+\frac{1}{2}\,K\,n_t\,m_{K_t}^3\right]\\
\sigma\,^{\rm disc}_{1/2}&\equiv&(3F-D)^2\,[\,m_{\eta_I}^3-3\,m_{\pi_I}^3\,]+\frac{a^2}{2}\sum _r\left[\,\delta_r'\,(\,K\,n_r+\frac{1}{2}\,{\cal S}\,^r_{abc}\,)\sum _{l \in L_r}R_l\,m_l^3\right]\\
\sigma\,^{\rm conn}_{3/2}&\equiv&\sum _t\left[\,(\,n_t+{\cal R}\,^t_{abc}\,)\,{\cal F}(m_{\pi_t})+\frac{1}{2}\,n_t\,{\cal F}(m_{K_t})\,\right]\\
\sigma\,^{\rm disc}_{3/2}&\equiv&a^2\,\sum _r\left[\,\delta'_r\,(\,2n_r+{\cal R}\,^r_{abc}\,)\,\sum _{l \in L_r}R_l\,{\cal F}(m_l)\,\right]\\
\sigma_{\rm val}&\equiv&-2(\alpha_M+\beta_M)\left[\,m_u+\frac{a^2}{14\lambda}\,(3\overline \Delta+4\,\overline {{\cal R}_{abc}\Delta})\,\right]\\
\sigma_{\rm sea}&\equiv&-2\sigma_M\left[\,2m_u+m_s+\frac{3a^2}{2\lambda}\overline \Delta\right],
\end{eqnarray*}
where overbars denote averages over the taste index $\alpha$,
and the valence and sea contributions in the tree-level terms have been separated:
${\rm Tree}(m_q) + {\rm tree}(a^2) = \sigma_{\rm val}+\sigma_{\rm sea}$.
We have verified that these contributions reduce to the
known result \cite{CHEN-SAVAGE} for the mass of the nucleon in the continuum limit.  
\begin{figure}
\begin{center}
\parbox[t]{0.45\textwidth}{
\includegraphics[width=0.45\textwidth]{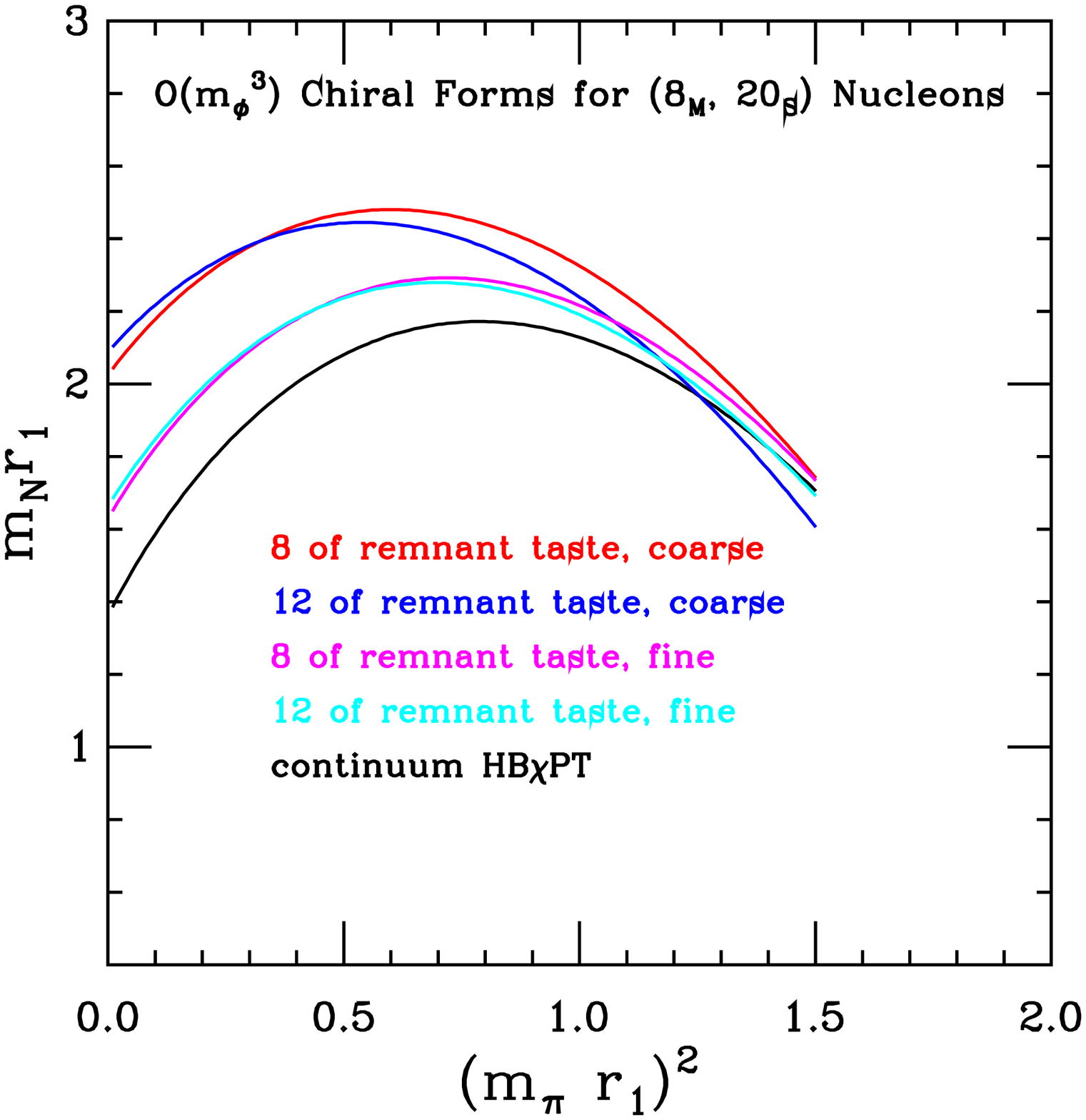}
}%
\hspace{0.3cm}
\parbox[t]{0.45\textwidth}{
\includegraphics[width=0.45\textwidth]{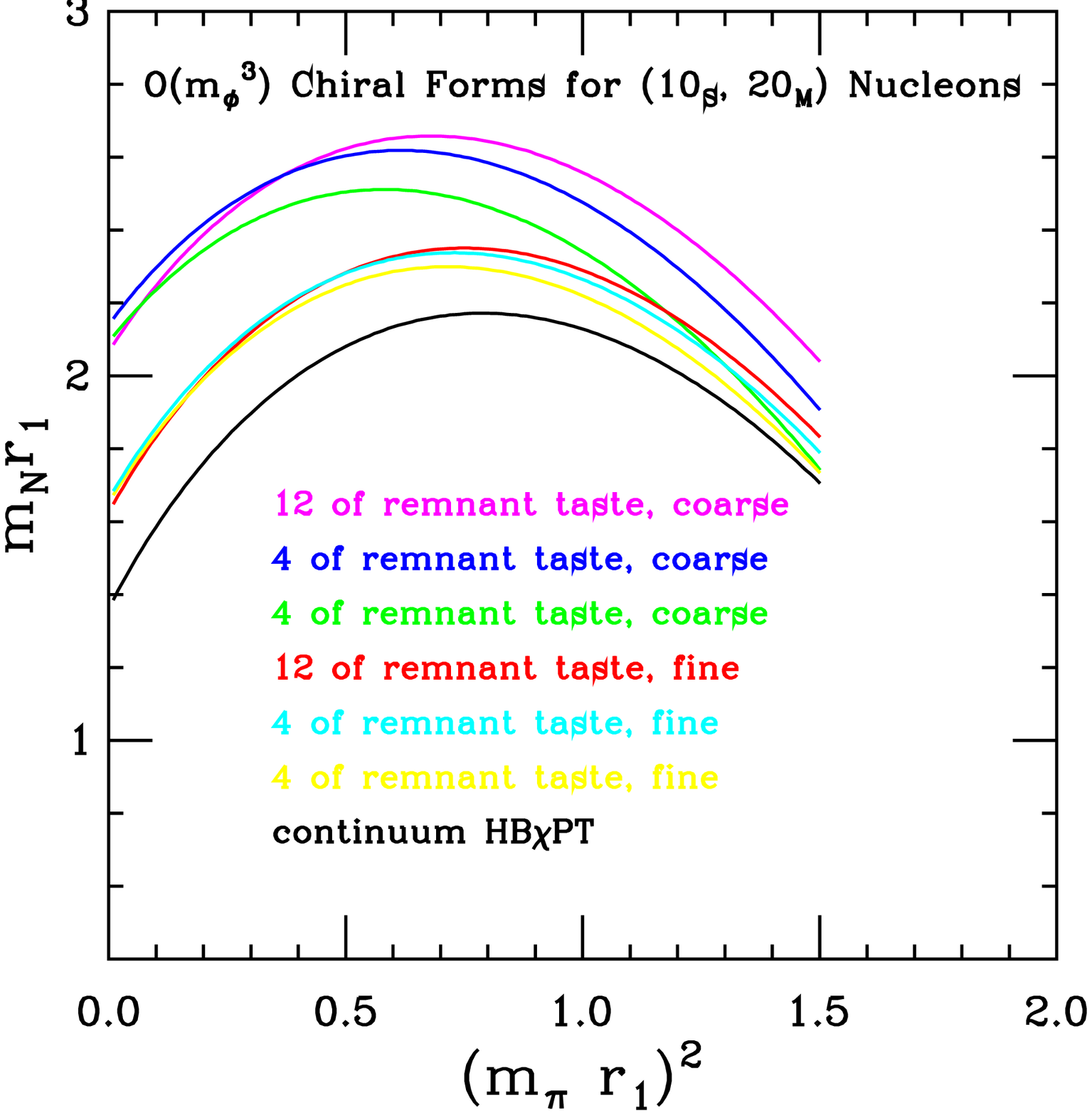}
}%
\end{center}
\caption{\label{FIG}The staggered chiral forms approach the continuum form as the lattice spacing decreases from the coarse lattice value to the fine.  To ${\cal O}(m_q^{3/2})$, the staggered nucleons are asymptotically proportional to $-m_{\phi}^3$ as $m_{\phi}\rightarrow \infty$.  }
\end{figure}
Using estimated values for the low-energy couplings and lattice parameters 
from recent simulations~\cite{4} 
yields 
a rough picture of the behavior of these chiral forms (Fig.~\ref{FIG}).  Because the contributions calculated to date are degenerate within irreps of the remnant taste $\Gamma_4\rtimes SO(4)_T$, there are two 
nondegenerate nucleons
 in the ${\bf (8_M,\ 20_S)}$ and three in the ${\bf (10_S,\ 20_M)}$.  Including all ${\rm Tree}(a^2)$ terms 
will break the remnant taste symmetry to the lattice group, 
splitting each of the {\bf 12}'s of Fig.~\ref{FIG} into two irreps of GTS:  ${\bf ({\rm 1\over 2},\ 12)}\rightarrow{\bf 8\oplus 16}$.

At small pion mass, the lattice data have mass and lattice spacing dependence qualitatively similar to that seen in Fig.~\ref{FIG}.  At larger mass higher order terms are clearly needed since the lattice data continue to increase for increasing $m_{\pi}$~\cite{11}.  Recent fits to continuum chiral forms \cite{12} suggest that the general trend of the data can be captured by including continuum corrections of ${\cal O}(m_q^2)$.  Therefore, it may be possible to neglect the effects of taste violations that are formally of ${\cal O}(m_q^2)$.  One may then be able to describe the relevant taste violations without a complete ${\cal O}(m_q^2)$ calculation in S$\chi$PT.

\section{\label{interp}Interpolating fields for ${\bf (10_S,\ 20_M)}$ nucleons}

The interpolating fields currently in use were constructed to transform as members of irreps of GTS \cite{17}.
Since the symmetry group of S$\chi$PT is the same as that of staggered QCD, the baryons of S$\chi$PT also transform within irreps of GTS.  Interpolating fields transforming within a given type of lattice irrep can create only those baryons of S$\chi$PT transforming within this same type of lattice irrep.

The staggered baryon interpolating fields 
in Ref.~\cite{17} were constructed for 1-flavor, 4-taste QCD, 
so they are symmetric under $SU(3)_F\subset SU(12)_{\rm val}$.  Because the only symmetric flavor 
representation
 in~(\ref{mix}) is the ${\bf 10_S}$, these operators can create only members of the ${\bf (10_S,\ 20_M)}$.  Interpolating fields transforming as members of ${\bf 8}$'s of GTS create baryons transforming as members of the three ${\bf 8}$'s in (\ref{20M to lat}), while interpolating fields transforming within ${\bf 16}$'s of GTS create baryons transforming within the ${\bf 16}$ in (\ref{20M to lat}).  

The three ${\bf 8}$'s of (\ref{20M to lat}) are split by taste violations, so interpolating fields transforming within ${\bf 8}$'s (cf. Table 3 of~\cite{17}) create spectra containing two excited states which are very near the ground state.  Unless S$\chi$PT provides predictions for these splittings, accounting for the contamination could be nontrivial.  In contrast, interpolating fields transforming within ${\bf 16}$'s of GTS create baryons transforming within the lone ${\bf 16}$ of (\ref{20M to lat}).  Fitting these correlators may therefore be cleaner, especially if one avoids operators that couple strongly to spin-$3\over 2$ states \cite{18}.

We have applied this analysis to the other $SU(3)_F\times SU(4)_T$ irreps in the decompositions of the ${\bf 572}$ and ${\bf 364}$.  However, creating multiple-flavor baryons such as those in the ${\bf (8_M,\ 20_S)}$ will require the construction of multiple-flavor interpolating fields.  We are 
extending the approach used to construct the 1-flavor operators to this case.

\section{\label{sum}Summary and outlook}

Using partially quenched, staggered HB$\chi$PT, we have calculated leading loops and analytic contributions to the masses of all 1-flavor staggered baryons in the ${\bf (10_S,\ 20_M)}$ and all 2-flavor staggered baryons in the ${\bf (8_M,\ 20_S)}$.  The resulting chiral forms fall into irreps of the remnant taste group and reduce to the corresponding forms of partially quenched HB$\chi$PT in the continuum limit.  The calculation of remaining analytic terms of ${\cal O}(a^2)$ is in progress.  The chiral forms for staggered nucleons in the ${\bf (10_S,\ 20_M)}$ are the ones needed to quantify taste violations in simulation data obtained with 1-flavor interpolating fields.  The data \cite{11} could probably be described by supplementing the leading loops and analytic terms with ${\cal O}(m_q^2)$ terms arising from next-to-leading order continuum operators \cite{WLOUD}.

The staggered baryon operators used in simulations to date often create spectra with excited states very close to the ground state.  For 1-flavor nucleons, operators transforming within ${\bf 16}$'s of GTS \cite{17} are an important exception.  To include isospin-breaking effects and calculate masses of other baryons, multiple-flavor operators are needed, so we are 
constructing such operators. 

We thank our colleagues in the MILC Collaboration for helpful discussions.  This work is supported by the U.S. Department of Energy under grant DE-FG02-91ER40628.


\begin{thebibliography}{99}
\bibitem{LEE-SHARPE}
  W.~J.~Lee and S.~R.~Sharpe, \emph{Partial flavor symmetry restoration for chiral staggered fermions}, \emph{Phys. Rev. D} {\bf 60} (1999) 114503 [{\tt hep-lat/9905023}].

  \bibitem{1} C.~Aubin and C.~Bernard, \emph{Pion and kaon masses in staggered chiral perturbation theory}, \emph{Phys. Rev. D} {\bf 68} (2003) 034014 [{\tt hep-lat/0304014}] and
  \emph{Pseudoscalar decay constants in staggered chiral perturbation theory}, \emph{Phys. Rev. D} {\bf 68} (2003) 074011 [{\tt hep-lat/0306026}];
C.\ Bernard, {\it Chiral logs in the presence of staggered
flavor symmetry breaking},
{\it Phys.\ Rev. D} {\bf 65} (2002) 054031.

  \bibitem{3} C.~Aubin and C.~Bernard, \emph{Staggered chiral perturbation theory with heavy-light mesons}, 
\emph{Nucl. Phys. B (Proc. Suppl.)} {\bf 140} (2005) 491-493 [{\tt hep-lat/0409027}].
  \bibitem{4} C.~Aubin {\it et al.} [MILC Collaboration], \emph{Light pseudoscalar decay constants, quark masses, and low energy constants from three-flavor lattice QCD}, \emph{Phys. Rev. D} {\bf 70} (2004) 114501 [{\tt hep-lat/0407028}].
  \bibitem{CHARMED} C.\ Aubin {\it et al.}, [Fermilab Lattice, MILC and HPQCD Collaborations],
{\it Semileptonic decays of $D$ mesons in three-flavor lattice QCD},
{\it Phys.\ Rev.\ Lett.\ }{\bf 94} (2005) 011601 [{\tt hep-ph/0408306}]
and
{\it Charmed meson decay constants in three flavor lattice QCD},
 {\it Phys.\ Rev.\ Lett.\ }{\bf 95} (2005) 122002 [{\tt hep-lat/0506030}]. 
  \bibitem{11} C.~Aubin {\it et al.}, \emph{Light hadrons with improved staggered quarks:  Approaching the continuum limit}, {Phys. Rev. D} {\bf 70} (2004) 094505 [{\tt hep-lat/0402030}]. 
  \bibitem{PAPER} Jon~A.~Bailey and C.~Bernard, in preparation.
  \bibitem{17} M.~F.~L.~Golterman and J.~Smit, \emph{Lattice baryons with staggered fermions}, \emph{Nucl. Phys. B} {\bf 255} (1985) 328-340.
  \bibitem{CHEN-SAVAGE} J.~W.~Chen and M.~J.~Savage, \emph{Baryons in partially quenched chiral perturbation theory}, \emph{Phys. Rev. D} {\bf 65} (2002) 094001 [{\tt hep-lat/0111050}].
  \bibitem{12} Doug~Toussaint, private communication.
  \bibitem{18} M.~Fukugita, N.~Ishizuka, H.~Mino, M.~Okawa and A.~Ukawa, \emph{Full QCD hadron spectroscopy with two flavors of dynamical Kogut-Susskind quarks on the lattice}, \emph{Phys. Rev. D} {\bf 47} (1993) 4739-4769.
  \bibitem{WLOUD} A.~Walker-Loud, \emph{Octet baryon masses in partially quenched chiral perturbation theory}, \emph{Nucl. Phys. A} {\bf 747} (2005) 476-507 [{\tt hep-lat/0405007}].
\end{thebibliography}
\end{document}